\documentclass{article}

\usepackage[titletoc,title]{appendix}

\usepackage{spconf}
\usepackage{cite}
\usepackage{enumerate}
\usepackage{graphicx}
\usepackage[cmex10]{amsmath} 
\usepackage{amssymb}
\usepackage{xspace}
\usepackage{subfigure}
\usepackage[lined,linesnumbered,ruled,commentsnumbered]{algorithm2e} 
\usepackage{colonequals}
\usepackage[mathscr]{euscript}
\usepackage{fixltx2e}    
\usepackage{amsthm}
\usepackage{mathrsfs}

\usepackage{xspace}

\usepackage{tikz}
\usepackage{pgfplots}

\usetikzlibrary{arrows,shapes,backgrounds,plotmarks,positioning}

\pgfplotsset{compat=newest}                         
\pgfplotsset{plot coordinates/math parser=false}
\newlength\figureheight
\newlength\figurewidth

\newtheorem{theorem}{Theorem}

\newtheorem{example}{Example}

\newtheorem{experiment}{Experiment}

\newcommand{\RZ}[1]{\mathsf{Z}_{#1}}

\newcommand{\RW}[1]{\mathsf{W}_{#1}}

\newcommand{\rv}{\mathbf{r}} 
\newcommand{\rvS}{\hat{\mathbf{r}}} 
\newcommand{\rS}{\hat{r}} 
\newcommand{\rvE}{\mathbf{r}^*} 
\newcommand{\rE}{r^*} 
\newcommand{\wv}{\mathbf{w}}         
\newcommand{\One}{\mathbf{1}}      

\newcommand{\RRCO}{\mathscr{R}_{\text{DC}}}

\newcommand{\Set}[1]{\{#1\}}

\newcommand{\Real}{\mathbb{R}}
\newcommand{\RealPP}{\mathbb{R}_{++}}    

\newcommand{\SPLIT}{\text{SPLIT}}

\newcommand{\NormTwo}[1]{\lVert #1 \rVert_2}

\newcommand{\SFM}{\text{SFM}}

\newcommand{\figref}[1]{Fig.~\ref{#1}}

\title{Fairness in Multiterminal Data Compression: A Splitting Method for The Egalitarian Solution}
\name{Ni~Ding$^{\star}$ \qquad David~Smith$^{\star}$ \qquad Parastoo~Sadeghi$^{\dagger}$ \qquad Thierry~Rakotoarivelo$^{\star}$}

\address{$^{\star}$ Data61, The Commonwealth Scientific and Industrial Research Organisation, Australia\\
    $^{\dagger}$Research School of Engineering, The Australian National University }
\begin{document}
%
\maketitle
\begin{abstract}
This paper proposes a novel splitting (SPLIT) algorithm to achieve fairness in the multiterminal lossless data compression problem. It finds the egalitarian solution in the Slepian-Wolf region and completes in strongly polynomial time. We show that the SPLIT algorithm adaptively updates the source coding rates to the optimal solution, while recursively splitting the terminal set, enabling parallel and distributed computation. The result of an experiment demonstrates a significant reduction in computation time by the parallel implementation when the number of terminals becomes large. The achieved egalitarian solution is also shown to be superior to the Shapley value in distributed networks, e.g., wireless sensor networks, in that it best balances the nodes' energy consumption and is far less computationally complex to obtain.
\end{abstract}
\section{Introduction}
\label{sec:intro}

There are many problems in signal processing involving source coding, e.g., the distributed source coding problem \cite{Xiong2004,Schonberg2004} that typically arises in the wireless sensor networks (WSNs), or the multimedia source coding problems in \cite{Girod2005,Aaron2004,Cheung2006}. They are also called data compression (DC) problems since they aim to use the minimum
code length to describe the correlated sources with the minimum information loss. It is well known that the achievable source coding rates for the lossless DC constitutes the Slepian-Wolf (SW) region \cite{SW1973,Cover1975}, which could become very large as the number of terminals/sources increases. Then, the problem is not only to reach the SW region, but also to find a solution with certain feature.

For a system where the terminals (also known as {\em users} or {\em nodes}) are equally privileged, e.g., a WSN, we always seek to attain fairness in the SW region. Several game-theoretic contributions in \cite{Madiman2008ISIT,Madiman2008,Ding2015Game} showed that the Shapley value \cite{Shapley1953Value} is one solution within the SW region. However, computing the Shapley value is intractable in large scale systems due to the exponentially growing complexity in the number of terminals. In addition, the Shapley value distributes the source coding rates based on the statistics of source data so that the terminals with more information are assigned higher rates. Such a method is less suitable for systems such as WSNs where balancing the battery energy consumption across all terminals is desired.

Thus, we consider another approach to attain fairness in the SW region through the egalitarian solution. We show that the egalitarian solution balances the source coding rate distribution, amongst terminals, in the SW region so that any cost proportional to the source coding rate is fairly allocated to the terminals. Finding the egalitarian solution is formulated using quadratic programming. We solve this problem with a novel splitting (SPLIT) algorithm that adaptively updates a cost allocation method until the egalitarian solution is attained. The SPLIT algorithm also recursively breaks each terminal set into two smaller subsets, allowing each of them to calculate its own optimal solution, both in a distributed and parallel manner. The result of an experiment shows that the completion time by adopting the parallel computation is significantly reduced as the number of terminals increases. In addition, the SPLIT algorithm completes in strongly polynomial time, which is much faster than the exponential time required to obtain the Shapley value.

\section{System Model}
\label{sec:System}

Let $\RZ{V}=(\RZ{i}:i\in V)$ be a vector of discrete random variables indexed by a finite set $V$. Assume that there is a distinct node/user $i$ privately observes an $n$-sequence $\RZ{i}^n$ of the terminal  $\RZ{i}$ that is i.i.d.\ generated according to the joint distribution $P_{\RZ{V}}$. The users are required to encode their observations in a way such that the source sequence $\RZ{V}^{n}$ can be reconstructed at the sink\footnote{The sink could be a data fusion/gathering node, e.g., a cluster header in a WSN.} from the source codes. This problem is called \textit{(multiterminal) data compression (DC)} or \textit{source coding} \cite{Cover2012ITBook}.

For the subsets $X,Y \subseteq V$, let $H(X)$ be the amount of randomness in $\RZ{X}$ measured by the Shannon entropy \cite{Cover2012ITBook} and $H(X|Y)=H(X \cup Y)-H(Y)$ be the conditional entropy of $\RZ{X}$ given $\RZ{Y}$.
Denote $\rv_V = (r_i \colon i \in V)$ a \textit{(source coding) rate vector} with each dimension $r_i$ denoting the expected code length at which user $i$ encodes his/her observations $\RZ{i}^{n}$. We call $\rv_V$ an \textit{achievable rate vector} if the source sequence $\RZ{V}^{n}$ can be recovered at the sink by letting the users encode $\RZ{V}^{n}$ at the rate designated by $\rv_V$.
Let $r$ be the \textit{sum-rate function} associated with the rate vector $\rv_V$ such that
$$ r(X)=\sum_{i\in X} r_i, \quad \forall X \subseteq V $$
with the convention $r(\emptyset)=0$.
An achievable rate vector must satisfy the Slepian-Wolf (SW) constraints \cite{SW1973,Cover1975}:\footnote{The interpretation of \eqref{eq:SWConstrs} is (a) the users in $X$ must reveal $H(X|V \setminus X)$, the information that is uniquely obtained by $X$, to the sink; (b) the users must reveal the total information $H(V)$ to the sink.}
\begin{equation} \label{eq:SWConstrs}
    \begin{aligned}
        r(X) & \geq H(X|V \setminus X), \quad \forall X \subseteq V, \\
        r(V) & = H(V).
    \end{aligned}
\end{equation}

For a subset $X \subseteq V$, consider the constraint $ r(X) \geq H(X|V \setminus X)$ in \eqref{eq:SWConstrs}. Since the sum-rate is restricted to $r(V) = H(V)$, there is necessarily an upper bound $ r(V \setminus X) = H(V) - r(X) \leq H(V \setminus X)$. Repeating the same conversion for all $X \subseteq V$, we have the SW region
$$ \RRCO(V,H) = \Set{ \rv_V \in P(H,\leq) \colon  r(V) = H(V) }, $$
where $ P(H,\leq) = \Set{ \rv_V \in \Real^{|V|} \colon  r(X) \leq H(X), \forall X \subseteq V }$ is the \textit{polyhedron} of the entropy function $H$ \cite[Section 2.2]{Fujishige2005}.
It is shown in \cite[Section 4.2]{FujishigePolyEntropy} that $H$ is submodular so that $\RRCO(V,H)$ coincides with a submodular base polyhedron. This crucial submodularity property underpins the optimality and efficiency of the SPLIT algorithm in Section~\ref{sec:SPLIT}.\footnote{In \cite{FujishigePolyEntropy}, $H$ is shown to be a polymatroid rank function, a subset of submodular functions.}

\section{Fairness: egalitarian vs. Shapley}
\label{sec:Fairness}
In general, the SW region $\RRCO(V,H)$ is not a singleton, i.e., there is usually more than one achievable rate vector. To choose a $\rv_V$ in $\RRCO(V,H)$ in a system where the users are equally privileged, e.g., sensors and mobile clients, a natural selection criterion is the fairness.
Based on the coalitional game formulation of the DC problem in \cite{Madiman2008ISIT,Madiman2008}, it is possible to reach the Shapley value \cite{Shapley1953Value} that attains fairness from a typical game-theoretic perspective.
Let $\sqcup$ denote the disjoint union. The Shapley value $\rvS_V$ with each dimension being \cite{Shapley1971Convex}
$$ \rS_i = \sum_{C \subseteq V \setminus \Set{i}} \frac{|C|! (|V| - |C| - 1)!}{|V|!} \big( H(C \sqcup \Set{i}) - H(C) \big)$$
allocates each user his/her expected marginal entropy.\footnote{$H(C \sqcup \Set{i}) - H(C)$ is the marginal randomness introduced by user $i$ to user subset $C$; By assuming each permutation of $V$ appears equiprobably, $\frac{|C|! (|V| - |C| - 1)!}{|V|!}$ quantifies the frequency that user $i$ will be assigned rate $H(C \sqcup \Set{i}) - H(C)$. The Shapley value $\rvS_V$ lies in $\RRCO(V,H)$ \cite{Shapley1971Convex}.}
Since the expected source coding length is determined by the entropy \cite{Cover2012ITBook}, the Shapley value is fair in that it assigns each user the cost he/she introduces to the DC problem by encoding his/her observations. However, this fair rate vector may not be the desired one in all cases. See the example below.

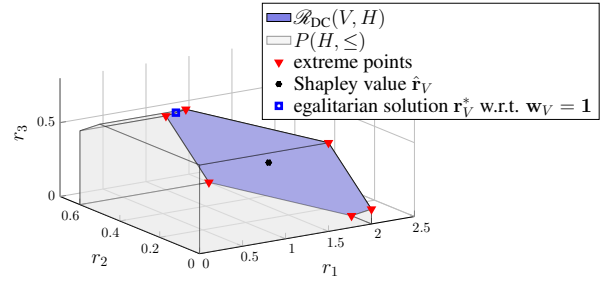
\begin{figure}[tpb]
	\centering
    \scalebox{0.53}{
%
%
%
\definecolor{mycolor1}{rgb}{0.5,0.5,0.9}%
\definecolor{mycolor2}{rgb}{1,0,1}%
\begin{tikzpicture}

\begin{axis}[%
width=3.5in,
height=2.1in,
view={-33}{30},
scale only axis,
xmin=0,
xmax=2.5,
xlabel={\Large $r_1$},
xmajorgrids,
ymin=0,
ymax=0.7,
ylabel={\Large $r_2$},
ymajorgrids,
zmin=0,
zmax=0.8,
zlabel={\Large $r_3$},
zmajorgrids,
axis x line*=bottom,
axis y line*=left,
axis z line*=left,
legend style={at={(0.57,1.18)},anchor=north west,draw=black,fill=white,legend cell align=left}
]

\addplot3[area legend,solid,fill=mycolor1,draw=black]
table[row sep=crcr]{
x y z\\
2 0.1 0 \\
2 0 0.1 \\
1.5 0 0.6 \\
1 0.5 0.6 \\
1 0.6 0.5 \\
1.5 0.6 0 \\
};
\addlegendentry{\Large $\RRCO(V,H)$};

\addplot3[area legend,solid,fill=white!90!black,opacity=4.000000e-01,draw=black]
table[row sep=crcr]{
x y z\\
2 0 0 \\
2 0.1 0 \\
2 0 0.1 \\
2 0 0 \\
};
\addlegendentry{\Large $P(H,\leq)$};

\addplot3[solid,fill=white!90!black,opacity=4.000000e-01,draw=black,forget plot]
table[row sep=crcr]{
x y z\\
2 0 0 \\
2 0 0.1 \\
1.5 0 0.6 \\
0 0 0.6 \\
0 0 0 \\
2 0 0 \\
};

\addplot3[solid,fill=white!90!black,opacity=4.000000e-01,draw=black,forget plot]
table[row sep=crcr]{
x y z\\
0 0 0.6 \\
1.5 0 0.6 \\
1 0.5 0.6 \\
0 0.5 0.6 \\
0 0 0.6 \\
};

\addplot3[solid,fill=white!90!black,opacity=4.000000e-01,draw=black,forget plot]
table[row sep=crcr]{
x y z\\
0 0.5 0.6 \\
1 0.5 0.6 \\
1 0.6 0.5 \\
0 0.6 0.5 \\
0 0.5 0.6 \\
};

\addplot3[solid,fill=white!90!black,opacity=4.000000e-01,draw=black,forget plot]
table[row sep=crcr]{
x y z\\
0 0.6 0.5 \\
1 0.6 0.5 \\
1.5 0.6 0 \\
0 0.6 0 \\
0 0.6 0.5 \\
};

\addplot3[solid,fill=white!90!black,opacity=4.000000e-01,draw=black,forget plot]
table[row sep=crcr]{
x y z\\
0 0 0 \\
0 0.6 0 \\
1.5 0.6 0 \\
2 0.1 0 \\
2 0 0 \\
0 0 0 \\
};

\addplot3[solid,fill=white!90!black,opacity=4.000000e-01,draw=black,forget plot]
table[row sep=crcr]{
x y z\\
0 0 0 \\
0 0 0.6 \\
0 0.5 0.6 \\
0 0.6 0.5 \\
0 0.6 0 \\
0 0 0 \\
};

\addplot3 [
color=red,
line width=2.0pt,
only marks,
mark=triangle,
mark options={solid,,rotate=180}]
table[row sep=crcr] {
2 0.1 0\\
2 0 0.1\\
1.5 0 0.6\\
1 0.5 0.6\\
1 0.6 0.5\\
1.5 0.6 0\\
};
\addlegendentry{\Large extreme points};

\addplot3 [
color=black,
line width=3.0pt,
only marks,
mark=asterisk,
mark options={solid}]
table[row sep=crcr] {
1.5 0.3 0.3\\
};
\addlegendentry{\Large Shapley value $\rvS_V$};

\addplot3 [
color=blue,
line width=2.0pt,
only marks,
mark=square,
mark options={solid}]
table[row sep=crcr] {
1 0.55 0.55\\
};
\addlegendentry{\Large egalitarian solution $\rvE_V$ w.r.t. $\wv_V = \One$};


\end{axis}
\end{tikzpicture}%
}
	\caption{The polyhedron $P(H,\leq)$ and the SW region $\RRCO(V,H)$ for the DC problem in Example~\ref{ex:main}. There are two fair rate vectors in $\RRCO(V,H)$: The Shapley value $\rvS_V = (\frac{3}{2},\frac{3}{10},\frac{3}{10})$ is the gravity center of the extreme points; The egalitarian solution $\rv_V = (1,\frac{11}{20},\frac{11}{20})$ is the minimum $\ell_2$-norm, i.e., the minimizer of $\min\Set{\NormTwo{\rv_V} \colon \rv_V \in \RRCO(V,H)}$.}
	\label{fig:demo}
\end{figure}

\begin{example} \label{ex:main}
The three users in $V = \Set{1,2,3}$ observe respectively
\begin{equation}
    \begin{aligned}
        \RZ{1} &= (\RW{a},\RW{b},\RW{c}), \\
        \RZ{2} &= (\RW{c},\RW{d}),   \\
        \RZ{3} &= (\RW{b},\RW{d}),
    \end{aligned} \nonumber
\end{equation}
where $\RW{j}$s for all $j \in \Set{a,\dotsc,d}$ are independent random bits with $H(\RW{a}) = 1$, $H(\RW{b}) = H(\RW{c}) = \frac{1}{2}$ and $H(\RW{d}) = \frac{1}{10}$. The SW region is shown in \figref{fig:demo}. We have the Shapley value $\rvS_V = (\frac{3}{2},\frac{3}{10},\frac{3}{10})$, in which user $1$ is assigned the most source coding rates $r_1 = \frac{3}{2}$ since he/she has the highest expected marginal entropy. However, consider the minimum $\ell_2$-norm $\rvE_V = \arg\min\Set{\NormTwo{\rv_V} \colon \rv_V \in \RRCO(V,H)} = (1,\frac{11}{20},\frac{11}{20})$. $\rvE_V$ allocates the rates more evenly in $\RRCO(V,H)$ than the Shapley value. For a WSN, $\rvE_V$ is better than $\rvS_V$ in that the maximum energy consumption in source coding $\max\Set{\rE_i \colon i \in V} < \max\Set{\rS_i \colon i \in V} $ is less and therefore the lifetime of WSN is prolonged.\footnote{The lifetime in a WSN is usually defined as the time to which the first sensor node runs out of battery power\cite{Dietrich2009}. Assume that the energy consumption is linearly proportional to the source coding rate, switching from the rate vector $(\frac{3}{2},\frac{3}{10},\frac{3}{10})$ to $(1,\frac{11}{20},\frac{11}{20})$ prolongs the lifetime by $\frac{1}{2}$. }
Here, $\rvE_V$ is called the egalitarian solution \cite{Dutta1989Egalitarian,Dutta1990EgaliConvex} in $\RRCO(V,H)$.\footnote{The advantage of the egalitarian solution in a peer-to-peer communication problem is also exemplified in \cite{Ding2016ISIT}.}
\end{example}

Consider a more general quadratic programming problem
\begin{equation} \label{eq:WDuttaRay}
    \min\Set{\sum_{i \in V} \frac{r_i^2}{w_i} \colon \rv_V \in \RRCO(V,H)},
\end{equation}
where $\wv_V \in \RealPP^{|V|}$ is a positive weight vector that could have some practical interpretation: $w_i$ could denote the quality of wireless channel at user $i$ or the remaining energy in sensor node $i$.
The minimizer $\rvE_V$ of \eqref{eq:WDuttaRay} is called the \textit{weighted, or generalized, egalitarian solution w.r.t. $\wv_V$}\cite{Hokari2002,Hokari2003}: It reduces to minimum $\ell_2$-norm when $\wv_V = \One = (1,\dotsc,1) \in \RealPP^{|V|}$.\footnote{The solution of \eqref{eq:WDuttaRay} also coincides with the min-max and max-min solutions \cite[Theorem 37]{Fujishige2009PP}: $\rvE_V = \arg\min \max_{i \in V} \Set{\frac{r_i}{w_i} \colon \rv_V \in \RRCO(V,H)} $ and $\rvE_V = \arg\max \min_{i \in V} \Set{\frac{r_i}{w_i} \colon \rv_V \in \RRCO(V,H)} $.}
For a given weight vector $\wv_V$, denote $w \colon 2^V \mapsto \RealPP$ the sum-weight function with $w(X) = \sum_{i \in X} w_i, \forall X \subseteq V$.

\section{Split Algorithm}
\label{sec:SPLIT}

To efficiently solve the problem~\eqref{eq:WDuttaRay}, we exploit the submodularity of the SW region $\RRCO(V,H)$ as identified in Section~\ref{sec:System}. The authors in \cite{Fujishige2009PP,Nagano2012Lex} showed that the minimizer of \eqref{eq:WDuttaRay} can be determined by recursively solving the submodular function minimization (SFM) problem, based on which, we propose the SPLIT algorithm in Algorithm~\ref{algo:Split}. Its optimality is given by the following theorem with the proof in Section~\ref{app:proof}.

\begin{theorem}\label{theo:OptSPLIT}
    The output $\rvE_V$ of the call $\SPLIT(V,H,\wv_V)$ is the minimizer of \eqref{eq:WDuttaRay}. \hfill\qed
\end{theorem}

	\begin{algorithm} [t]
	\label{algo:Split}
	\small
	\SetAlgoLined
    \SetKwInOut{Input}{input}\SetKwInOut{Output}{output}
	\SetKwRepeat{Repeat}{repeat}{until}
    \SetKwIF{If}{ElseIf}{Else}{if}{then}{else if}{else}{endif}
    \SetKw{Return}{return}
    \Input{a user subset $C$, an oracle that returns the value of $f(X)$ for $X \subseteq C$ and a positive weight vector $\wv_C$}
	\Output{$\rvE_C = \arg\min\Set{\sum_{i \in C} \frac{r_i^2}{w_i} \colon \rv_C \in \RRCO(C,f)}$}
	\BlankLine
	\Begin{
        $\lambda \leftarrow \frac{ f(C) }{w(C)}$\;
        get the maximal minimizer $\hat{X}$ of
        \begin{equation} \label{eq:SPLITmin}
            \min \Set{f(X)-\lambda w(X) \colon X \subseteq C};
        \end{equation}
        \eIf{$\hat{X} = C$}{
            $\rvE_C \leftarrow \lambda \wv_C$;
            }{
                $\rvE_{\hat{X}} = \SPLIT(\hat{X},f,\wv_{\hat{X}})$\;
                $\rvE_{C \setminus \hat{X}} \leftarrow \frac{f(\hat{X})}{w(\hat{X})} \wv_{C \setminus \hat{X}}$\;
                $\rvE_{C \setminus \hat{X}} \leftarrow \rvE_{C \setminus \hat{X}} + \SPLIT(C \setminus \hat{X},g,\wv_{C \setminus \hat{X}})$, where
                $$ g(X) = f(X \sqcup \hat{X}) - f(\hat{X}) (\frac{w(X)}{w(\hat{X})} +1), \forall X \subseteq C \setminus \hat{X};$$
                $\rvE_C = \rvE_{\hat{X}} \oplus \rvE_{C \setminus \hat{X}}$;
            }
        return $\rvE_C$\;
        }	
	\caption{split algorithm (SPLIT)}
	\end{algorithm}

\begin{example} \label{ex:Split}
    For $\wv_V = (3,1,3)$, we apply the SPLIT algorithm to the system in Example~\ref{ex:main}. By calling $\SPLIT(V,H,\wv_V)$, we have $\lambda = \frac{3}{10}$ and $\hat{X} = \Set{3}$ being the maximal minimizer of $\min \Set{H(X)-\lambda w(X) \colon X \subseteq V}$. Since $\hat{X} \neq V$, we call run $\SPLIT(\Set{3},H,w_3)$ and get $\rE_3 = \frac{3}{5}$ returned. We set $\rvE_{\Set{1,2}} = \frac{H(\Set{3})}{w_3} \wv_{\Set{1,2}} = (\frac{1}{5}w_1,\frac{1}{5}w_2) = (\frac{3}{5},\frac{1}{5})$ and call $\SPLIT(\Set{1,2},g,\wv_{\Set{1,2}})$ where $g(X) = H(X \sqcup \Set{3}) - H(\Set{3}) (\frac{w(X)}{w_3} +1) $ for all $X \subseteq \Set{1,2}$. We get output $\rvE_{\Set{1,2}} = (0.525,0.175)$, which is added to the current rates so that we have $\rvE_{\Set{1,2}} = (\frac{3}{5},\frac{1}{5}) + (0.525,0.175) = (\frac{9}{8},\frac{3}{8})$. Finally, we have $\rvE_V = \rvE_{\Set{1,2}} \oplus \rE_3 = (\frac{9}{8},\frac{3}{8},\frac{3}{5})$ at the output which is the egalitarian solution w.r.t. $\wv_V = (3,1,3)$ in $\RRCO(V,H)$.

    We then run $\SPLIT(V,H,\wv_V)$ for $\wv_V = \One$, where $\lambda = \frac{3}{10}$ and $\hat{X} = \Set{2,3}$ is the maximal minimizer of $\min \Set{H(X)-\lambda |X| \colon X \subseteq V}$. Since $\hat{X} \neq V$, we call $\SPLIT(\Set{2,3},H,\wv_{\Set{2,3}})$ to get $\rvE_{\Set{2,3}} = (\frac{11}{20},\frac{11}{20})$ and $\SPLIT(\Set{1},g,\wv_{\Set{1}})$ to get $\rE_1 = 1 $ so that $\rvE_V = \rE_1 \oplus \rvE_{\Set{2,3}} = (1,\frac{11}{20},\frac{11}{20})$ is the egalitarian solution in \figref{fig:demo}.
\end{example}

\subsection{Complexity}

For a submodular function $f$, we call $\min \Set{ f(X) \colon X \subseteq V}$ a \textit{submodular function minimization (SFM) problem} of \textit{size} $|V|$. We denote $O(\SFM(|V|))$ the complexity  for solving this SFM problem, which is strongly polynomial \cite[Chapter VI]{Fujishige2005}. While the SFM algorithms in \cite{Orlin2009SFM,Iwata2007SFM,IFF2001,Fujishige2011MinNorm} vary in computation complexity, the exact completion time of a SFM algorithm depends on its size $|V|$.\footnote{For example, the complexity of the SFM algorithm proposed in \cite{Orlin2009SFM} is in the order of $|V|^5$.}
It is easy to see that, due to the submodularity of $H$, the problem \eqref{eq:SPLITmin} in each recursion of the SPLIT algorithm is a SFM of size $|C| \leq |V|$.\footnote{The minimizers of a SFM problem form a lattice, where the maximal and minimal subsets exist, and the maximal minimizer can be determined at the same time when the SFM is solved \cite{Fujishige2005}. }
Since the number of recursions is no greater than $2|V| - 1$ \cite[Theorem 9]{Nagano2012Lex}, the overall complexity is upper bounded by $O(|V| \cdot \SFM(|V|))$.

\subsection{Adaptive and Distributed Implementation}
\begin{figure}[tpb]
	\centering
    \scalebox{0.53}{
%
%
%
\definecolor{mycolor1}{rgb}{0.5,0.5,0.9}%
\definecolor{mycolor2}{rgb}{1,0,1}%
\begin{tikzpicture}

\begin{axis}[%
width=3.5in,
height=2.1in,
view={-33}{30},
scale only axis,
xmin=0,
xmax=2.5,
xlabel={\Large $r_1$},
xmajorgrids,
ymin=-0.01,
ymax=0.7,
ylabel={\Large $r_2$},
ymajorgrids,
zmin=0,
zmax=0.8,
zlabel={\Large $r_3$},
zmajorgrids,
axis x line*=bottom,
axis y line*=left,
axis z line*=left,
legend style={at={(0.47,1.23)},anchor=north west,draw=black,fill=white,legend cell align=left}
]

\addplot3[area legend,solid,fill=mycolor1,draw=black,forget plot]
table[row sep=crcr]{
x y z\\
2 0.1 0 \\
2 0 0.1 \\
1.5 0 0.6 \\
1 0.5 0.6 \\
1 0.6 0.5 \\
1.5 0.6 0 \\
};

\addplot3[area legend,solid,fill=white!90!black,opacity=4.000000e-01,draw=black,forget plot]
table[row sep=crcr]{
x y z\\
2 0 0 \\
2 0.1 0 \\
2 0 0.1 \\
2 0 0 \\
};

\addplot3[solid,fill=white!90!black,opacity=4.000000e-01,draw=black,forget plot]
table[row sep=crcr]{
x y z\\
2 0 0 \\
2 0 0.1 \\
1.5 0 0.6 \\
0 0 0.6 \\
0 0 0 \\
2 0 0 \\
};

\addplot3[solid,fill=white!90!black,opacity=4.000000e-01,draw=black,forget plot]
table[row sep=crcr]{
x y z\\
0 0 0.6 \\
1.5 0 0.6 \\
1 0.5 0.6 \\
0 0.5 0.6 \\
0 0 0.6 \\
};

\addplot3[solid,fill=white!90!black,opacity=4.000000e-01,draw=black,forget plot]
table[row sep=crcr]{
x y z\\
0 0.5 0.6 \\
1 0.5 0.6 \\
1 0.6 0.5 \\
0 0.6 0.5 \\
0 0.5 0.6 \\
};

\addplot3[solid,fill=white!90!black,opacity=4.000000e-01,draw=black,forget plot]
table[row sep=crcr]{
x y z\\
0 0.6 0.5 \\
1 0.6 0.5 \\
1.5 0.6 0 \\
0 0.6 0 \\
0 0.6 0.5 \\
};

\addplot3[solid,fill=white!90!black,opacity=4.000000e-01,draw=black,forget plot]
table[row sep=crcr]{
x y z\\
0 0 0 \\
0 0.6 0 \\
1.5 0.6 0 \\
2 0.1 0 \\
2 0 0 \\
0 0 0 \\
};

\addplot3[solid,fill=white!90!black,opacity=4.000000e-01,draw=black,forget plot]
table[row sep=crcr]{
x y z\\
0 0 0 \\
0 0 0.6 \\
0 0.5 0.6 \\
0 0.6 0.5 \\
0 0.6 0 \\
0 0 0 \\
};

\addplot3 [
color=red,
line width=2.0pt,
only marks,
mark=triangle,
mark options={solid,,rotate=180},forget plot]
table[row sep=crcr] {
2 0.1 0\\
2 0 0.1\\
1.5 0 0.6\\
1 0.5 0.6\\
1 0.6 0.5\\
1.5 0.6 0\\
};

\addplot3 [
color=black,
line width=3.0pt,
only marks,
mark=asterisk,
mark options={solid}]
table[row sep=crcr] {
1.5 0.3 0.3\\
};
\addlegendentry{\Large Shapley value $\rvS_V$};

\addplot3 [
color=blue,
line width=2.0pt,
only marks,
mark=square,
mark options={solid}]
table[row sep=crcr] {
1 0.55 0.55\\
};
\addlegendentry{\Large egalitarian solution $\rvE_V$ w.r.t. $\wv_V = \One$};

\addplot3 [
->,
color=blue,
dashed,
mark=star,
mark options={solid,,rotate=180},
line width=2.0pt]
table[row sep=crcr] {
0 0 0\\
0.55 0 0\\
1 0.55 0.55\\
};
\addlegendentry{\Large update path to $\rvE_V$ w.r.t. $\wv_V = \One$};

\addplot3 [
color=mycolor2,
line width=3.0pt,
only marks,
mark=star,
mark options={solid}]
table[row sep=crcr] {
1.125 0.375 0.6\\
};
\addlegendentry{\Large egalitarian solution $\rvE_V$ w.r.t. $\wv_V = (3,1,3)$};

\addplot3 [
->,
color=mycolor2,
dashed,
mark=star,
mark options={solid,,rotate=180},
line width=2.0pt]
table[row sep=crcr] {
0 0 0\\
0.6 0.2 0\\
1.125 0.375 0.6\\
};
\addlegendentry{\Large update path to $\rvE_V$ w.r.t. $\wv_V = (3,1,3)$};

\end{axis}
\end{tikzpicture}%
}
	\caption{The rate adaptation resulted in the SPLIT algorithm in Example~\ref{ex:Split}:
$(0,0,0) \rightarrow (\frac{11}{20},0,0) \rightarrow (1,\frac{11}{20},\frac{11}{20}) $ to the egalitarian solution w.r.t. $\wv_V = \One$; $(0,0,0) \rightarrow (\frac{3}{5},\frac{1}{5},0) \rightarrow (\frac{9}{8},\frac{3}{8},\frac{3}{5}) $ to the egalitarian solution w.r.t. $\wv_V = (3,1,3)$; }
	\label{fig:path}
\end{figure}
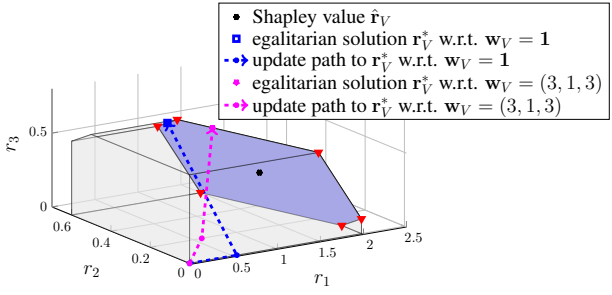

Steps 7 and 8 in the SPLIT algorithm indicate an adaptive rate update method: In step 7, since $\rvE_{C \setminus \hat{X}} \geq \frac{f(\hat{X})}{w(\hat{X})} \wv_{C \setminus \hat{X}}$,\footnote{This is based on the proof of Theorem~\ref{theo:OptSPLIT} in Section~\ref{app:proof}.}
we first assign $\frac{f(\hat{X})}{w(\hat{X})} \wv_{C \setminus \hat{X}}$ to $\rvE_{C \setminus \hat{X}}$ and determine the remaining rates by calling $\SPLIT(C \setminus \hat{X},g,\wv_{C \setminus \hat{X}})$ in step 8. It is easy to see that $\rvE_V \in P(H,\leq)$ after each execution of step 7. Therefore, the SPLIT algorithm adaptively updates the rate vector in the polyhedron $P(H,\leq)$ until it finally reaches $\rvE_V$. See the examples in \figref{fig:path}.

In addition, steps 6 to 8 in the SPLIT algorithm can be implemented in a distributed manner: The set $C$ is split into two disjoint subsets $\hat{X}$ and $C \setminus \hat{X}$; $\rvE_{\hat{X}}$ and $\rvE_{C \setminus \hat{X}}$ are independently obtained and the computation can be done in parallel. In \figref{fig:distribute}, we show how to obtain the egalitarian solution $\rvE_V = (\frac{9}{8},\frac{3}{8},\frac{3}{5})$ in Example~\ref{ex:Split} in a distributed manner.
Since the SPLIT algorithm recursively splits $C$ into two, it usually results in a tree diagram where the splitting is from top to bottom. If parallel computation is allowed, the completion time in each split is dominated by the maximum size of SFM $\max \Set{|C \setminus \hat{X}|,|\hat{X}|}$.

\begin{figure}[tpb]
	\centering
    \scalebox{0.65}{
%
%
\begin{tikzpicture}

\begin{axis}[%
width=3.6in,
height=1.8in,
scale only axis,
xmin=3,
xmax=80,
xlabel={\Large $|V|$, number of terminals/users},
xmajorgrids,
ymin=0,
ymax=250,
ylabel={\Large mean size of SFM},
ymajorgrids,
legend style={at={(0.02,0.97)},anchor=north west,draw=black,fill=white,legend cell align=left}
]

\addplot [
color=blue,
solid,
line width = 1,
mark=triangle,
mark options={solid},
]
table[row sep=crcr]{
3 6.014\\
4 8.784\\
5 11.808\\
6 14.717\\
7 17.151\\
8 20.167\\
9 23.022\\
10 26.46\\
11 29.975\\
12 32.835\\
13 34.731\\
14 37.984\\
15 42.137\\
16 44.513\\
17 46.688\\
18 51.085\\
19 54.417\\
20 56.882\\
21 62.661\\
22 62.687\\
23 64.714\\
24 69.827\\
25 72.386\\
26 75.41\\
27 77.772\\
28 83.473\\
29 86.549\\
30 87.644\\
31 91.251\\
32 92.408\\
33 95.412\\
34 101.669\\
35 101.461\\
36 108.966\\
37 106.613\\
38 114.773\\
39 116.288\\
40 118.619\\
41 121.406\\
42 124.824\\
43 129.002\\
44 132.231\\
45 139.081\\
46 141.862\\
47 141.275\\
48 142.993\\
49 145.807\\
50 156.219\\
51 153.53\\
52 160.356\\
53 170.102\\
54 165.882\\
55 165.546\\
56 167.842\\
57 171.131\\
58 173.777\\
59 187.165\\
60 183.108\\
61 185.655\\
62 192.931\\
63 190.912\\
64 192.103\\
65 196.833\\
66 201.419\\
67 201.446\\
68 206.834\\
69 209.65\\
70 215.651\\
71 220.375\\
72 223.706\\
73 226.191\\
74 222.396\\
75 232.709\\
76 241.933\\
77 240.828\\
78 241.026\\
79 237.426\\
80 247.145\\
};
\addlegendentry{\large nonparallel implementation};

\addplot [
color=red,
solid,
line width = 1,
mark=o,
mark options={solid},
]
table[row sep=crcr]{
3 4.855\\
4 6.872\\
5 9.159\\
6 11.34\\
7 13.198\\
8 15.334\\
9 17.416\\
10 19.864\\
11 22.274\\
12 24.216\\
13 25.858\\
14 28.097\\
15 30.621\\
16 32.688\\
17 34.317\\
18 37.431\\
19 39.414\\
20 41.249\\
21 45.127\\
22 45.395\\
23 46.745\\
24 50.334\\
25 51.793\\
26 54.414\\
27 56.124\\
28 59.338\\
29 61.607\\
30 62.025\\
31 64.804\\
32 66.363\\
33 68.252\\
34 72.439\\
35 72.291\\
36 76.434\\
37 76.138\\
38 80.765\\
39 82.335\\
40 84.611\\
41 86.217\\
42 89.002\\
43 91.569\\
44 92.609\\
45 97.099\\
46 99.178\\
47 100.522\\
48 101.173\\
49 103.074\\
50 109.743\\
51 107.753\\
52 112.168\\
53 117.817\\
54 115.617\\
55 116.305\\
56 118.155\\
57 120.657\\
58 121.985\\
59 130.155\\
60 128.077\\
61 129.905\\
62 136.021\\
63 133.165\\
64 132.16\\
65 137.296\\
66 140.037\\
67 140.965\\
68 145.838\\
69 146.124\\
70 149.647\\
71 153.939\\
72 154.727\\
73 157.39\\
74 155.368\\
75 160.267\\
76 166.676\\
77 165.723\\
78 167.69\\
79 166.465\\
80 169.597\\
};
\addlegendentry{\large parallel implementation};

\end{axis}
\end{tikzpicture}%
}
	\caption{The result in Experiment~\ref{exp:Parallel}: The average sum-size and maximum size of SFM for the nonparallel and parallel implementations, respectively, of the SPLIT algorithm.}
	\label{fig:Complexity}
\end{figure}
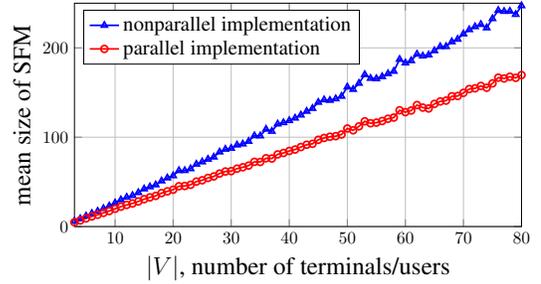

\begin{experiment} \label{exp:Parallel}
Let $|V|$ vary from $3$ to $80$. For each value of $|V|$, the following procedure is repeated for $100$ times: (a) randomly generate $\RZ{V}$; (b) run the SPLIT algorithm and get the sum-size of SFM $|C \setminus \hat{X}| + |\hat{X}|$ and maximum size of SFM $\max \Set{|C \setminus \hat{X}|,|\hat{X}|}$ and sum them up over recursions, which indicate the completion time of centralized/nonparallel and parallel implementations, respectively. We average the sum-size and maximum size of SFM over repetitions and show them in \figref{fig:Complexity}. It can be seen that the parallel implementation is much faster when terminal/user set $V$ becomes large.
\end{experiment}

\section{Conclusion}
We proposed the SPLIT algorithm for determining the egalitarian solution, a fair source coding rate vector, in the SW region for the multiterminal lossless DC problem. We proved that the SPLIT algorithm gradually updates a rate vector to the egalitarian solution in strongly polynomial time. We showed how it recursively splits the terminal/user set to allow distributed and parallel computation. We confirmed experimentally that the parallel implementation greatly reduces the completion time as the number of terminals/users grows. We also showed that, compared to the Shapley value, the egalitarian solution balances energy consumed according to the source coding rates, making it a more suitable fairness metric in distributed systems.
Our future works include the implementation of SPLIT in a WSN deployment for precision agriculture, and the study of the fairness in the DC problem in a given data gathering tree.

\begin{figure}[tpb]
	\centering
    \scalebox{0.65}{\begin{tikzpicture}

\draw [draw=black] (-1.6,5.4) rectangle (1.6,4.6);
\node at (0,5) {\Large $1 \quad\quad 2 \quad\quad 3$};

\draw [draw=black] (-4,4) rectangle (-2,3.2);
\node at (-3,3.6) {\Large $1 \quad\quad 2$};
\draw [->] (-1.6,5)--(-3,5)--(-3,4);
\node at (-3,2.8) {\large \textcolor{blue}{ $r_1^*=\frac{9}{8} \quad r_2^*=\frac{3}{8}$}};

\draw [draw=black] (2.6,4) rectangle (3.4,3.2);
\node at (3,3.6) {\Large $3$};
\draw [->] (1.6,5)--(3,5)--(3,4);
\node at (3,2.8) {\large \textcolor{blue}{ $r_3^*=\frac{3}{5}$}};

\end{tikzpicture} }
	\caption{Distributed implementation of the SPLIT algorithm in Example~\ref{ex:Split} for obtaining the egalitarian solution $\rvE_V = (\frac{9}{8},\frac{3}{8},\frac{3}{5})$ w.r.t. $\wv_V = (3,1,3)$: The user set $V = \Set{1,2,3}$ is broken into two subsets $\Set{1,2}$ and $\Set{3}$ so that $\rvE_{\Set{1,2}}$ and $\rE_3$ can be determined in parallel.}
	\label{fig:distribute}
\end{figure}
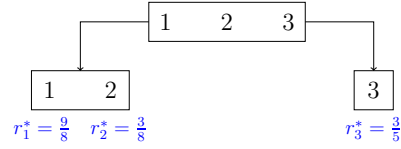

\section{Proof of Theorem 1}
\label{app:proof}

It is shown in \cite[Section 3.4]{Fujishige2009PP} that problem~\eqref{eq:WDuttaRay} can be solved by considering the problem $\min \Set{ H(X) - \lambda w(X) \colon X \subseteq V }$. The minimizer for all $\lambda \in \Real$ is fully characterized by $p \leq |V|$ critical points $ 0 = \lambda_0 < \lambda_1 < \dotsc < \lambda_p $ and corresponding set chain $\emptyset = S_0 \subset S_1 \subset \dotsc \subset S_p = V$, where $S_j$ is the maximal minimizer of $\min \big\{ H(X) - \lambda_j w(X) \colon X \subseteq V \big\} $.\footnote{Also, for all $j \in \Set{1,\dotsc,p}$, $S_{j-1}$ is the unique minimizer of $\min \big\{ f(X) - \lambda w(X) \colon X \subseteq V \big\} $ if $\lambda_{j-1} < \lambda < \lambda_{j}$.}
The minimizer $\rvE_V$ of \ref{eq:WDuttaRay} can be obtained by
$ \rE_i = \lambda_j w_i, \forall i \in S_j \setminus S_{j-1}, j \in \Set{1,\dotsc,p}. $

The property of $\lambda_j$ and $S_j$ is derived in \cite[Lemmas 5 and 6]{Nagano2012Lex}: For $j,j' \in \Set{0,\dotsc,p} \colon j > j'$ and $\lambda  = \phi(S_j,S_{j'}) = \frac{H(S_j) - H(S_{j'})}{w(S_j \setminus S_{j'})}$, $\lambda  = \lambda_j$ if $j = j' + 1$; $\lambda_{j'+1} < \lambda < \lambda_j$ if $j > j' + 1$. It apparently suggests an recursive method to determine $\rvE_V$: For $\lambda = \phi(S_j,S_{j'})$, determine the maximal minimizer $\hat{X}$ of $\min \Set{ f(X) - \lambda w(X) \colon X \subseteq V }$; If $\hat{X} = S_{j}$, set $\rE_i = \lambda w_i$ for all $i \in S_j \setminus S_{j'}$ and terminate recursion; Otherwise, repeat the same procedure for $\bar{\lambda} = \phi(S_j,\hat{X})$ and $\underline{\lambda} = \phi(\hat{X},S_{j'})$. By initiate $S_j = V$, $S_{j'} = \emptyset$ and $f = H$, this procedure determines $\rvE_V$.
For $S_j$ and $S_{j'}$, define $ g(X) = f(X \sqcup S_{j'}) - f(S_{j'}) (\frac{w(X)}{w(S_{j'})} +1), \forall X \subseteq V' = S_j \setminus S_{j'} $ and $\lambda' = \frac{g(V')}{w(V')}$. It can be shown that $\lambda = \phi(S_j,S_{j'})  = \lambda' + \frac{f(S_{j'})}{w(S_{j'})}$ and $g(X) - \lambda' w(X) = f(X \sqcup S_{j'}) - f(S_{j'}) - \lambda w(X)$ for all $X \subseteq V'$. Therefore, $\arg\min \Set{ g(X) \colon X \subseteq V' } \sqcup S_{j'} = \arg\min \Set{ f(X) - \lambda w(X) \colon X \subseteq V }$ and $\lambda \wv_{S_j \setminus S_{j'}} = \lambda' \wv_{S_j \setminus S_{j'}} + \frac{f(S_{j'})}{w(S_{j'})} \wv_{S_j \setminus S_{j'}}$. Theorem holds.    \qed

\vfill\pagebreak

\bibliographystyle{IEEEbib}
\bibliography{CGBIB}

\end{document}